\documentclass[aps,prb,twocolumn]{revtex4} 

\usepackage{graphicx}
\usepackage{dcolumn}
\usepackage{bm}
\usepackage{amsmath} 

\newcommand{\comment}[1]{}

\begin{document}

\title{Power Series for the Quantum Statistical Mechanics Probability
\\ with Results for the Second Virial Coefficient of Helium}



\author{Phil Attard}
\affiliation{ {\tt phil.attard1@gmail.com} }


\begin{abstract}
A power series for the Wigner-Kirkwood pair commutation function
for quantum statistical mechanics in classical phase space
is given with terms automatically generated by recursion.
The calculated second virial coefficient agrees with the measured values
of helium for temperatures greater than 65\,K.
Prospects for a general quantum Monte Carlo algorithm are discussed.
\end{abstract}


\maketitle

%
\section{Introduction}
\setcounter{equation}{0} \setcounter{subsubsection}{0}
\renewcommand{\theequation}{\arabic{section}.\arabic{equation}}
%

Although a many-particle open quantum system
may be represented exactly as a configuration in classical phase space,
the proximal impediment to viable computer treatments
is the evaluation of the Wigner-Kirkwood commutation function.
The latter was
originally formulated by Wigner (1932),
and later in phase space by Kirkwood (1933).
The latter gave a high temperature expansion,
several variants of which have been developed and studied
by the present author (Attard 2018b, 2021).
These expansions are unwieldy to derive,
complicated to implement,
and slowly converging in practice,
and to date they have had limited success in simulations
of condensed matter such as the noble gases
(Attard 2016b, 2017, 2018b, 2021, 2025a).

Nevertheless there are several reasons
to continue to seek viable formulations of the commutation function.
From the conceptual point of view it is useful to understand
how the classical world arises from its quantum foundations,
which includes the dynamics of particles in classical phase space
as determined by Schr\"odinger's equation (Attard 2025b).
Similarly wave function symmetrization,
and the dependent phenomenon of Bose-Einstein condensation,
are most easily understood and computed in classical phase space
(Attard 2025a, 2025g).
Finally,
path integral quantum Monte Carlo,
which is the main competing computer simulation method,
is much more computationally demanding
(Allen and Tildesley 1987),
which restricts it  to small system sizes
on the order of 64 atoms (Ceperley 1995),
compared to systems with 5,000 atoms that are routinely simulated
using the commutation function (Attard 2018b, 2021, 2025a).

This paper presents a new approach to the commutation function.
The differential equation for its temperature derivative
is approximated by pair terms.
The pair commutation function is expanded
as a series in powers of the temperature and three components
of the separation and momentum difference.
This gives a simple recursion relation
for the coefficients of the expansion
that is suitable for automated implementation on a computer.
An expression for the second virial coefficient is also derived,
and numerical results using the Lennard-Jones pair potential
are obtained.
When restricted to regions with imaginary phase magnitude less than $\pi/2$,
as proposed in previous work (Attard 2025h),
the results for helium are in good agreement with measured values
for temperatures above 65\,K.

The present power series expansion
is by far the most promising approach to the commutation function
that has been tested to date.
Prospects for using it as the basis of a more general computer
simulation algorithm are discussed in the conclusion.

%
\section{Fundamentals} \label{Sec:Analysis}
\setcounter{equation}{0} \setcounter{subsubsection}{0}
\renewcommand{\theequation}{\arabic{section}.\arabic{equation}}
%

\subsection{Phase Space Weight}

For a subsystem of  $N$ identical bosons,
a point in classical phase space is
${\bf \Gamma} = \{{\bf q},{\bf p}\}$,
and its conjugate with reversed momenta is
${\bf \Gamma}^\dag = \{{\bf q},-{\bf p}\}$.
Here the position configuration is
${\bf q} = \{ {\bf q}_1,{\bf q}_2,\ldots, {\bf q}_N\}$
and the momentum configuration is
${\bf p} = \{ {\bf p}_1,{\bf p}_2,\ldots, {\bf p}_N\}$,
where the position of boson $j$ is
${\bf q}_j = \{ {q}_{jx},{q}_{jy},{q}_{jz}\}$,
and its momentum is
${\bf p}_j = \{ {p}_{jx},{p}_{jy},{p}_{jz}\}$.
In many cases the momentum may belong to the continuum;
but for Bose-Einstein condensation and superfluidity quantized momentum
appears to be necessary (Attard 2025a, 2025b, 2025g).

For a canonical equilibrium system of temperature $T$ and volume $V$,
the complex phase space probability density is
(Attard 2018b, 2021)
\begin{equation}
\wp({\bf \Gamma}) =
\frac{
e^{-\beta {\cal H}({\bf \Gamma})}
e^{W({\bf \Gamma})} \eta({\bf \Gamma})
 }{ N! h^{3N}Z } \Theta\big({\pi}/{2}-|W_{\rm i}({\bf \Gamma})|\big) .
\end{equation}
Here Planck's constant is $h=6.63\times10^{-34}$\,J\,s
and the inverse temperature is $\beta = 1/k_{\rm B}T$,
$k_{\rm B} = 1.38 \times 10^{-23}$\,J/K
being Boltzmann's constant.
The partition function, $Z(N,V,T)$,
normalizes the probability and its logarithm gives the total entropy.
The classical Hamiltonian is
${\cal H}({\bf \Gamma}) = {\cal K}({\bf p}) + U({\bf q})$,
where ${\cal K}({\bf p}) = p^2/2m = \sum_{j=1}^N p_j^2/2m$
is the kinetic energy,
$m$ being the mass of a boson.
Usually the potential energy consists of central pair potentials,
$U({\bf q}) = \sum_{j<k}^N u(q_{jk})$;
one can write $u_{jk}  \equiv u(q_{jk})$.


The Wigner-Kirkwood (Kirkwood 1933, Wigner 1932)
commutation function $W({\bf \Gamma})$ is defined by
(Attard 2018b, 2021)
\begin{equation} \label{Eq:Wdefn}
e^{-\beta {\cal H}({\bf \Gamma})}
e^{W({\bf \Gamma})}
=
e^{{\bf p}\cdot{\bf q}/{\rm i} \hbar}
e^{-\beta \hat{\cal H}({\bf q})}
e^{-{\bf p}\cdot{\bf q}/{\rm i} \hbar} .
\end{equation}
In previous notation
this corresponds to $\omega_p =e^{W_p}$ (Attard 2018b Eq.~(2.6)).
For the quantum statistical weight one can use either $\omega_p \eta_q$
or else $\omega_q \eta_p$,
with $\omega_q^* = \omega_p$ and $\eta_q^* = \eta_p$.

On the right hand side of the definition of the commutation function
the Hamiltonian operator appears,
$ \hat{\cal H}({\bf q}) =  \hat{\cal K}({\bf q}) +  U({\bf q})$.
The Fourier factors are unnormalized, unsymmetrized momentum eigenfunctions.
The non-zero commutator $[\hat{\cal K}, U] \ne 0$
is what makes $W \ne 0$,
and so it reflects the Heisenberg uncertainty relation.

In the phase space weight appears the Heaviside step function $\Theta$
that is zero unless the imaginary part
of the commutation function has magnitude less than $\pi/2$,
in which case it equals unity.
This condition was proposed in previous work (Attard 2025h).
It was justified as the continuation of the restriction
to the positive phase space weight that holds at high temperatures.
Since in general the Wigner-Kirkwood commutation function
is extensive, phase space points giving values outside this range
belong to a neighborhood
with highly oscillatory and therefore canceling weight,
and the system can never cross such a zero-weighted boundary
as the temperature is lowered.
It will be shown in the results below for the second virial coefficient
that imposing this condition gives full agreement with the measured data,
whereas not imposing the constraint gives noticeable error.



The symmetrization function
is the ratio of unpermuted to permuted momentum eigenfunctions
summed over all permutations  (Attard 2018b, 2021),
\begin{equation}
\eta({\bf \Gamma})
=
\sum_{\hat{\rm P}}
e^{-[{\bf p}-\hat{\rm P}{\bf p}]\cdot{\bf q}/{\rm i}\hbar} .
\end{equation}
This is for bosons and it corresponds to $\eta_q$  (Attard 2018b Eq.~(2.4)).

Previous calculations of the symmetrization contribution
to the second virial coefficient
show it to be negligible for $^4$He for $T \agt 10$\,K
(Attard 2026b).
Hence the symmetrization function will be neglected in the rest
of the present paper.

\comment{ 
That part of the grand potential due to symmetrization
is given by the average of this,
\begin{eqnarray}
e^{-\beta \Omega_{\rm sym}}
& = &
\langle \eta({\bf \Gamma}) \rangle_W
\nonumber \\ & = &
e^{ \langle \stackrel{\circ}{\eta}({\bf \Gamma}) \rangle_W } .
\end{eqnarray}
In the second equality,
which is believed to be exact in the thermodynamic limit
(Attard 2018b \S~III\,B\,1),
the sum over single permutation loops is
$\stackrel{\circ}{\eta}\!\!({\bf \Gamma}) =
\sum_{l=2}^\infty \eta^{(l)}({\bf \Gamma})$,
where the $l$-loop symmetrization function is
\begin{equation}
\eta^{(l)}({\bf \Gamma})
=
\sum_{j_1,\ldots,j_l}^N\!\!\!'\; \prod_{k=1}^{l}
e^{ - {\bf p}_{j_k} \cdot {\bf q}_{j_k,j_{k+1}} /{\rm i}\hbar}
 , \quad j_{l+1} \equiv j_1 .
\end{equation}
The sum is over the unique directed cyclic permutations
of all subsets of $l$-bosons.
The $l$-loop grand potential is
(Attard 2018b, 2021)
\begin{equation}
-\beta \Omega_W^{(l)}
=
\langle \eta^{(l)}(\Gamma) \rangle_W .
\end{equation}
(The monomer grand potential is just the logarithm
of the partition function in the absence of the symmetrization function.)
Several tricks have been found to facilitate
the computation of the  $l$-loop symmetrization function
(Attard 2021 \S5.4.2).
On the high temperature side of the $\lambda$-transition,
we expect that only position loops with consecutive particles
separated by less than about the thermal wavelength will contribute.

} 

\subsection{Pair Commutation Function}

The inverse temperature derivative
of the defining equation for the commutation function,
Eq.~(\ref{Eq:Wdefn}), yields
(Attard 2021 Eq.~(8.57) or Eq.~(11.4))
\begin{eqnarray} \label{Eq:dWp/dB}
\frac{\partial W}{\partial \beta }
& = &
\frac{\mathrm{i}\hbar}{m} {\bf p} \cdot \nabla (W-\beta U)
+ \frac{\hbar^2}{2m}
\nabla^2 (W-\beta U)
\nonumber \\ && \mbox{ }
+ \frac{\hbar^2}{2m}
\nabla (W-\beta U) \cdot \nabla (W-\beta U) .
\end{eqnarray}

As mentioned, we consider a system with interaction potential
the sum of central pair potentials, $U({\bf q}) = \sum_{j<k} u_{jk}$.
The  quadratic term here is non-linear,
and it makes the commutation function a many-body function.
In this work we are primarily interested in the second virial coefficient,
and so we need only consider the pair contribution to this
\begin{equation}
W({\bf q},{\bf p})
\approx
\sum_{j<k}^N w^{(2)}({\bf q}_{jk},{\bf p}_{jk})
=
\frac{1}{2} \sum_{j,k}\!^{(k \ne j)} w_{jk}.
\end{equation}
We shall discuss the utility of this approximation
for the general case of computer simulations of condensed matter
in the conclusion.

Define $\tilde w \equiv w - \beta u$,
which,
apart from the kinetic energy,
is the dimensionless exponent for the classical phase space weight.
Retaining only the pair part of the non-linear term,
the temperature derivative equation becomes
(Attard 2021 Eq.~(11.10))
\begin{eqnarray}
\frac{\partial \tilde w_{jk}}{\partial \beta }
& = &
- u_{jk} +
\frac{\mathrm{i}\hbar}{m}
{\bf p}_{jk} \cdot \nabla_j \tilde w_{jk}
+ \frac{\hbar^2}{m} \nabla_j^2  \tilde w_{j k}
\nonumber \\ &&  \mbox{ }
+ \frac{\hbar^2}{m}
\nabla_j \tilde w_{j k} \cdot  \nabla_j \tilde w_{j k} .
\end{eqnarray}
Here and above
the symmetry $\tilde w_{j k} = \tilde w_{kj}$ has been exploited.
The high temperature limit gives the classical result,
$\tilde w_{j k} \sim -\beta u_{jk}$, $\beta \rightarrow 0$.

For a central pair potential we have
\begin{equation}
\nabla_a u_{ab}
= -\nabla_b u_{ab}
=  \frac{u_{ab}'}{q_{ab}} {\bf q}_{ab} ,
\end{equation}
and
\begin{equation}
\nabla_a^2 u_{ab}
= \nabla_b^2 u_{ab}
= u_{ab}'' + 2 \frac{u_{ab}'}{q_{ab}}.
\end{equation}
Note that  $\tilde w^{(2)}$
is not a central pair function.

\subsection{Power Series}

For two atoms $j$ and $k$ we write the momentum difference as
${\bf p} = {\bf p}_j - {\bf p}_k$
and the position difference ${\bf q} = {\bf q}_j - {\bf q}_k$.
From the symmetry of the problem we need only consider
the $z$-component of momentum difference, $p_z$,
the $z$-component of position difference, $q_z$,
and the separation $q$,
and we can write $\tilde w(p_z,q_z,q)$.

We define the dimensionless position $\tilde {\bf q} \equiv {\bf q}/\sigma$,
where $\sigma$ is the molecular diameter,
and the dimensionless momentum $\tilde {\bf p} \equiv {\bf p}\sigma /\hbar$,
and the dimensionless inverse temperature, $\tilde \beta = \beta \varepsilon$,
where $\varepsilon$ is related to the well-depth of the pair potential.

The power series to be used is
\begin{equation}
\tilde w(\tilde p_z,\tilde q_z,\tilde q)
= \sum_{j,k,l,n} w_{jkln}
\tilde \beta^j ({\rm i}\tilde p_z)^k \tilde q_z^l \tilde q^{-n} ,
\end{equation}
with the integers $\{j,k,l,n\}$ being non-negative.
We have to treat $\tilde q_x$ and $\tilde q_z$
as the independent position variables
in taking the gradients,
and at the end convert $q_x^2 = q^2 -q_z^2$.

One has the gradients
\begin{equation}
\tilde \nabla_x \left( \tilde q_z^l \tilde q^{-n} \right)
=
-n \tilde q_z^l \tilde q^{-n-2}  \tilde q_x
\end{equation}
and
\begin{eqnarray}
\tilde \nabla_z \left( \tilde q_z^l \tilde q^{-n} \right)
& = &
l \tilde q_z^{l-1} \tilde q^{-n}
- n \tilde q_z^{l+1} \tilde q^{-n-2} .
\end{eqnarray}
And similarly for the higher order gradients.

Inserting these into the temperature derivative
and equating the same powers on both sides
we obtain the recursion relation
for $j \ge 1$, 
\begin{eqnarray}
\lefteqn{
(j+1) w_{j+1,k,l,n}
} \nonumber \\
& = &
\frac{\hbar^2}{m\varepsilon\sigma^{2}}
\left\{ (l+1)  w_{j,k-1,l+1,n} - (n-2) w_{j,k-1,l-1,n-2} \right\}
\nonumber \\ &  & \quad
+
\frac{\hbar^2}{m\varepsilon\sigma^{2}}
\big[ (l+2)(l+1) w_{j,k,l+2,n}
\nonumber \\ &  & \quad
+ (n-2)(n - 2l - 2) w_{j,k,l,n-2}
\nonumber \\ &  & \quad
- 2(n-4)(n-2) w_{j,k,l-2,n-4}  \big]
\nonumber \\ &  & \quad
+
\frac{\hbar^2}{m\varepsilon\sigma^{2}}
\sum_{j',k',l',n'} w_{j',k',l',n'}
\nonumber \\ &  & \quad \times
\Big\{
n'(n-n'-2) w_{j-j',k-k',l-l',n-n'-2}
\nonumber \\ && \quad
- n'(n-n') w_{j-j',k-k',l-l'-2,n-n'}
\nonumber \\ && \quad
+ l'(l-l'+2) w_{j-j',k-k',l-l'+2,n-n'}
\nonumber \\ && \quad
- (l'(n-n'-2)+(l-l')n') w_{j-j',k-k',l-l',n-n'-2}
\nonumber \\ && \quad
+ n'(n-n'-4)w_{j-j',k-k',l-l'-2,n-n'-4}
\Big\} .
\end{eqnarray}
Although this ten-line equation might appear a little daunting for a human,
for a computer it is simply par for the course.

In this work we use the Lennard-Jones potential,
in which case the starting values for $j=1$ are
\begin{equation} \label{Eq:w1-LJ}
w_{1kln} =
-4 [ \delta_{n,12} - \delta_{n,6} ] \delta_{k,0} \delta_{l,0} .
\end{equation}
From this the computer can successively generate all coefficients
up to some maximum.
Coefficients with indeces outside the range are zero.
For Lennard-Jones helium-4,
the parameter multiplying the right hand side is
${\hbar^2}/{m\varepsilon\sigma^{2}} =0.18$
(the corresponding temperature where this multiplied by $\tilde \beta$
equals unity is 1.9\,K),
which suggests that the higher order coefficients
quickly  become negligible.
For the results presented below in a typical case
there were 19 non-zero coefficients.

It is also possible to use a more sophisticated pair potential,
such as the Hartree-Fock dispersion-B2 pair potential (Aziz 1992),
by starting with a fit of
$\sum_n w_{1kln} \tilde \beta \tilde q^{-n} $
to $-\beta  u(q)$.
This is further discussed in the results section.

\subsection{Virial Pressure}

The thermodynamic pressure for a canonical equilibrium system is given
by the volume derivative of the Helmholtz free energy,
which is proportional to the logarithm of the partition function
(Attard 2002, 2012),
\begin{equation} \label{Eq:BpV}
\beta p = \frac{\partial \ln Z}{\partial V}
= \frac{1}{3L^2 Z }  \frac{\partial  Z}{\partial L} .
\end{equation}

The position integrals are over the cubic volume $V=L^3$.
We scale all lengths by $L$, ${\bf q} = L{\bf x}$,
with $x_{j\alpha} \in [-1/2,1/2]$.
Hence $\partial {\bf q} /\partial L = L^{-1}  {\bf q}$.
The quantized momenta have spacing $\Delta_p = 2\pi\hbar/L$
and so at constant quantum state,
$\partial {\bf p} /\partial L = -L^{-1}  {\bf p}$.
In this case products such as ${\bf p}\cdot {\bf q}$
are independent of $L$.
This includes products with permuted configurations,
${\bf p}\cdot \hat{\rm P} {\bf q}$,
which means that the symmetrization function is independent of $L$.

The virial derivative for the pressure gives
\begin{eqnarray}
\beta p & = &
\frac{1}{3V}
\left\langle {\bf \Gamma}^\dag
\cdot \nabla_{\bf \Gamma}
[ -\beta {\cal K}({\bf p}) - \beta U({\bf q}) + W({\bf \Gamma}) \right\rangle
\nonumber \\ & = &
\frac{1}{3V} \left\langle 2 \beta {\cal K}({\bf p})  \right\rangle
+ \left\langle {\bf \Gamma}^\dag
\cdot \nabla_{\bf \Gamma} \tilde w({\bf \Gamma}) \right\rangle .
\end{eqnarray}

The virial expansion for the pressure is
$\beta p = B_1 \rho + B_2 \rho^2 + \ldots$,
where the number density is $\rho = N/V$
(Attard 2002 \S~8.1, Pathria 1972 \S~9.2).
For fixed number $N$,
the contribution to the pressure
from the first virial coefficient is proportional to $V^{-1}$,
and that from the second virial coefficient is proportional to $V^{-2}$.

The partition function for two particles is
\begin{eqnarray}
Z
& = &
\frac{1}{2! h^6}
\int {\rm d}{\bf \Gamma}_1 \, {\rm d}{\bf \Gamma}_2\;
e^{-\beta [p_1^2 + p_2^2]/2m} e^{\tilde w(p_{21,z},q_{21,z},q_{21})}
\nonumber \\ & = &
\frac{2\pi V}{ \Lambda^6 2!}
\int_{-\infty}^{\infty}  {\rm d} p_z
\int_L {\rm d}{q}\, q^2
\int_{-1}^1 {\rm d}x\;
\omega_{p}(p_{z})
e^{\tilde w(p_{z},q_z,q)},
\nonumber \\
\end{eqnarray}
with $q_z = qx$.
Here and below the thermal wavelength is
$\Lambda = \sqrt{2\pi \beta \hbar^2  /m}$.
The second equality follows by invoking
the relative momentum and separation,
and with the relative momentum density being
$\omega_{p}(p_{z})
= ({\Lambda }/{h\sqrt{2}}) e^{-\beta  p_z ^2/4m } $,
with $\int {\rm d}p_z \,\omega_{p}(p_{z})
= 1$.

\subsubsection{Kinetic Energy Contribution}

Let us define the analogue of the Mayer-$f$ function,
\begin{equation}
\tilde f(p_{z},q_z,q)
\equiv
e^{\tilde w(p_{z},q_z,q)} - 1.
\end{equation}
Because $\tilde w$ is a function
of the pair potential $u(q)$ and its derivatives,
$\tilde w \to 0$ as $q\to\infty$,
and so $\tilde f$ also  goes to zero for large separations $q$.
With this we can write the  two-particle partition function as
$Z = Z^{\rm id} + Z_f$, where
\begin{eqnarray}
Z_{\rm id} & = &
\frac{1}{ 2! h^6}
\int {\rm d}{\bf \Gamma}_1 \, {\rm d}{\bf \Gamma}_2\;
e^{-\beta [p_1^2 + p_2^2]/2m}
\nonumber \\ & = &
\frac{V^2}{2\Lambda^6} ,
\end{eqnarray}
and
\begin{eqnarray}
Z_f & = &
\frac{1}{ 2! h^6}
\int {\rm d}{\bf \Gamma}_1 \, {\rm d}{\bf \Gamma}_2\;
e^{-\beta [p_1^2 + p_2^2]/2m}
\tilde f(p_{z},q_z,q)
 \\ & = & \nonumber
\frac{2\pi V}{ \Lambda^6 2!}
\int_{-\infty}^{\infty}  {\rm d} p_z
\int_0^\infty {\rm d}{q}\, q^2
\int_{-1}^1 {\rm d}x\;
\omega_{p}(p_{z})
\tilde f(p_{z},q_z,q) .
\end{eqnarray}
We have   $Z^{\rm id} \propto V^2$
and $Z_f \propto V$.
Note that $q_z = qx$,
and since $\tilde f(p_{z},q_z,q) \to 0$ $q\to \infty$,
we can drop the dependence on $V=L^3$
and extend the integral over the pair separation $q$ to infinity.

In view of these the kinetic energy contribution to the pressure is
\begin{eqnarray}
\left\langle 2 \beta {\cal K}({\bf p})  \right\rangle
& = &
\frac{2 \beta}{2m}   4 \frac{m}{\beta}
+
\frac{2 \beta}{2m}
\left\langle  p_{1,z}^2 +  p_{2,z}^2 \right\rangle
\nonumber \\ & = &
4
+
\frac{\beta}{2 m}
\left\langle  p_+^2 + p_-^2 \right\rangle
\nonumber \\ & = &
5
+ \frac{\beta}{2 m} \left\langle p_z^2 \right\rangle .
\end{eqnarray}
Recall that $p_z \equiv p_- \equiv  p_{21,z} \equiv p_{2,z}-p_{1,z}$.
It is only the final average that depends upon
$\tilde w(p_z,q_z,q)$.
This may be written
\begin{eqnarray}
\left\langle p_z^2  \right\rangle
& = &
\frac{1}{2h^6 Z}\!
\int \! {\rm d}{\bf \Gamma}_1 \, {\rm d}{\bf \Gamma}_2\,
e^{-\beta [p_1^2 + p_2^2]/2m} e^{\tilde w(p_{21,z},q_{21,z},q_{21})}
p_z^2
\nonumber \\ & = &
\frac{1}{Z^{\rm id} + Z_f}
\frac{2\pi V}{2! \Lambda^6 }
\int_{-\infty}^{\infty}  {\rm d} p_z
\int_L {\rm d}{q}\, q^2
\int_{-1}^1 {\rm d}x\;
\nonumber \\ && \quad \times
\omega_{p}(p_{z})  [\tilde f(p_{z},q_z,q)+1] p_z^2
\nonumber \\ & = &
\frac{Z^{\rm id}}{Z^{\rm id} + Z_f}
\left[ \left\langle \tilde f(p_{z},q_z,q)  p_z^2 \right\rangle_{\rm id}
+ \left\langle \tilde   p_z^2 \right\rangle_{\rm id} \right]
\nonumber \\ & = &
\left\langle \tilde   p_z^2 \right\rangle_{\rm id}
+
\left\langle \tilde f(p_{z},q_z,q)  p_z^2 \right\rangle_{\rm id}
\nonumber \\ && \quad
- \left\langle \tilde   p_z^2 \right\rangle_{\rm id}
\left\langle   \tilde f(p_{z},q_z,q) \right\rangle_{\rm id}
+ {\cal O}(V^{-2}) .
\end{eqnarray}
We have $\langle p_z^2 \rangle_{\rm id} = 2m/\beta$,
$Z^{\rm id} = {V^2}/{2\Lambda^6}$
and, for $k=0,2$,
\begin{eqnarray}
\lefteqn{
\left\langle p_z^k \tilde f(p_z,q_z,q) \right\rangle_{\rm id}
}  \\
& = &
\frac{2\pi V}{2! \Lambda^6  Z_{\rm id}}\!
\int_{-\infty}^\infty \!\! {\rm d} p_z   e^{ -\beta p_{z}^2/4m}\!
\int_0^\infty \!\! {\rm d}q \, q^2\!\!
\int_{-1}^{1} \!\!{\rm d}x \,
\tilde f(p_z,q_z,q) p_z^k
\nonumber \\ & = &
\frac{2\pi }{V} \!
\int_{-\infty}^\infty \!{\rm d} p_z \, e^{ -\beta p_{z}^2/4m}\!
\int_0^\infty \!{\rm d}q \, q^2 \!
\int_{-1}^{1} \!{\rm d}x \,
\tilde f(p_z,q_z,q) p_z^k  .\nonumber
\end{eqnarray}
(Actually, $\tilde f$ should be restricted to the primary branch
that is the positive real half of the complex plane.)

Putting these together, the contribution to the pressure from
the virial derivative of the kinetic energy is
\begin{eqnarray}
\left\langle 2 \beta {\cal K}({\bf p})  \right\rangle
& = &
6
+ \frac{\beta}{2m}
\left\langle p_z^2 \tilde f(p_z,q_z,q) \right\rangle_{\rm id}
 \\ && \quad
-  \frac{\beta}{2m} \left\langle p_z^2 \right\rangle_{\rm id}
\left\langle \tilde f(p_z,q_z,q) \right\rangle_{\rm id}
+ {\cal O}(V^{-2}) .\nonumber
\end{eqnarray}
The first term is the classical ideal gas contribution.
Hence
\begin{equation}
\beta p^{\rm id}
=
\frac{1}{3V}
\left\langle 2 \beta {\cal K}({\bf p})  \right\rangle_{\rm id}
= \frac{6}{3V}
=\frac{2}{V} = \rho .
\end{equation}
This gives the first virial coefficient, $B_1=1$.
(In the virial expansion, the contribution to the pressure
from the first virial coefficient is proportional to $V^{-1}$,
and that from the second virial coefficient is proportional to $V^{-2}$.)

The contribution to the second virial coefficient
from the average of the kinetic energy comes from the remainder
\begin{eqnarray}
\lefteqn{
B_2^{({\cal K})}
=
\frac{\beta p^{(2,{\cal K})}}{ \rho^2}
}  \\
& = &
\frac{V^2}{6V}
\left[
\left\langle 2 \beta {\cal K}({\bf p})  \right\rangle
-
\left\langle 2 \beta {\cal K}({\bf p})  \right\rangle_{\rm id}
\right]
\nonumber \\ & = &
\frac{\beta V}{12 m}
\left\{
\left\langle p_z^2 \tilde f(p_z,q_z,q) \right\rangle_{\rm id}
-  \left\langle p_z^2 \right\rangle_{\rm id}
\left\langle \tilde f(p_z,q_z,q) \right\rangle_{\rm id}
\right\}
\nonumber \\ & = &
\frac{2\pi\beta}{12m}
\int_{-\infty}^\infty {\rm d} p_z
\int_0^\infty {\rm d}{q} \, q^2
\int_{-1}^{1}  {\rm d} x\;
\frac{ e^{-\beta p_z^2/4m} }{\sqrt{4\pi\beta/m}}
\nonumber \\ && \quad \times
\tilde f(p_{z},q_z,q)
[ p_z^2 - 2m \beta^{-1} ]
\Theta(\pi/2-|\tilde w_{\rm i}|). \nonumber
\end{eqnarray}
Here  $\rho^2 = N(N-1)/V^2 = 2/V^2$ for $N=2$.
Note that the integrand consists of two-body terms
($p_z$ is the $z$-component of the momentum difference)
and that this expression is completely analogous to
the second virial coefficient derived
from the pair commutation function exponent below,
with $2\beta p_z^2/4m $ replacing $v^{\rm int}$.
The embraced term is ${\cal O}(V^{-1})$,
and so the right hand side is independent of volume.
This has the form of a fluctuation about the average
and can be expected to be small.
Note that since $\tilde f$ is complex
we have included the Heaviside step function
to keep the imaginary exponent
on the primary branch that is the positive real half of the complex plane.


\subsubsection{Potential Energy Contribution}

The remaining part of the pressure is given by
\begin{eqnarray}
\beta p^{(\tilde w)} & = &
\frac{1}{3V}
\left\langle {\bf \Gamma}^\dag
\cdot \nabla_{\bf \Gamma} \tilde w({\bf \Gamma}) \right\rangle
 \\ & = &
\frac{1/3V}{2! h^6 Z}
\int {\rm d}{\bf \Gamma}_1 \, {\rm d}{\bf \Gamma}_2\;
e^{-\beta [p_1^2 + p_2^2]/2m} e^{\tilde w}\,
{\bf \Gamma}^\dag
\cdot \nabla_{\bf \Gamma} \tilde w({\bf \Gamma}) .\nonumber
\end{eqnarray}
Because $\tilde w$ is short-ranged,
the explicit double integral is ${\cal O}(V)$.
To leading order,
we can replace $Z = Z^{\rm id} + Z_f \Rightarrow Z^{\rm id}$,
since $Z^{\rm id} = {\cal O}(V^2)$ and  $Z_f={\cal O}(V)$.
This gives a denominator that is ${\cal O}(V^3)$.
Using   $\rho^2 \Rightarrow 2/V^2$,
the contribution to the second virial coefficient from
$\tilde w$ is
\begin{equation}
B_2^{(\tilde w)}
=
\frac{V^2}{2} \frac{1}{3V} \left\langle
 e^{\tilde w}\,
 {\bf \Gamma}^\dag \cdot \nabla_{\bf \Gamma} \tilde w({\bf \Gamma})
\right\rangle_{\rm id} .
\end{equation}

We have
\begin{eqnarray}
\lefteqn{
v^{\rm int}(p_z,q_z,q)
} \nonumber \\
& \equiv &
{\bf \Gamma}^\dag
\cdot \nabla_{\bf \Gamma} \tilde w({\bf \Gamma})
\nonumber \\ & = &
-{\bf p}_1 \cdot \nabla_{{\bf p}_1} \tilde w
-{\bf p}_2 \cdot \nabla_{{\bf p}_2} \tilde w
+
{\bf q}_1 \cdot \nabla_{{\bf q}_1} \tilde w
+
{\bf q}_2 \cdot \nabla_{{\bf q}_2} \tilde w
\nonumber \\ & = &
- p_{21,z} \nabla_{p_z} \tilde w 
+ q_{21,x} \nabla_{q_x} \tilde w 
+ q_{21,z} \nabla_{q_z} \tilde w. 
\end{eqnarray}
Recall the  power series
$\tilde w(\tilde p_z,\tilde q_z,\tilde q)
= \sum_{j,k,l,n} w_{jkln}
\tilde \beta^j ({\rm i}\tilde p_z)^k \tilde q_z^l \tilde q^{-n}$
and the dimensionless variables $\tilde {\bf q} \equiv {\bf q}/\sigma$,
$\tilde {\bf p} \equiv {\bf p}\sigma /\hbar$,
and  $\tilde \beta = \beta \varepsilon$.
The three contributions to $v^{\rm int} $ are
\begin{equation}
- \tilde p_{z}
\tilde \nabla_{p_z} \tilde w(\tilde p_{z},\tilde q_{z},\tilde q)
=
- \sum_{j,k,l,n}  k w_{jkln}
\tilde \beta^j ({\rm i}\tilde p_z)^k \tilde q_z^l \tilde q^{-n} ,
\end{equation}
\begin{eqnarray}
\lefteqn{
\tilde q_{x} \tilde \nabla_{q_x} \tilde w(\tilde p_{z},\tilde q_{z},\tilde q)
} \nonumber \\
& = &
\sum_{j,k,l,n}  -n w_{jkln}
\tilde \beta^j ({\rm i}\tilde p_z)^k
\frac{\tilde q_z^l}{\tilde q^{n+1} } \frac{\tilde q_x}{\tilde q} \tilde q_x
\nonumber \\ & = &
\sum_{j,k,l,n}   w_{jkln} \tilde \beta^j ({\rm i}\tilde p_z)^k
\left[ n\frac{\tilde q_z^{l+2}}{\tilde q^{n+2}}
- n\frac{\tilde q_z^l}{\tilde q^{n}} \right],
\end{eqnarray}
and
\begin{eqnarray}
\lefteqn{
\tilde q_{z} \tilde \nabla_{q_z} \tilde w(\tilde p_{z},\tilde q_{z},\tilde q)
} \nonumber \\
& = &
\sum_{j,k,l,n} w_{jkln} \tilde \beta^j ({\rm i}\tilde p_z)^k
\left[ l  \frac{\tilde q_z^{l}}{\tilde q^n }
- n  \frac{\tilde q_z^{l+2}}{\tilde q^{n+2} } \right] .
\end{eqnarray}

With these the contribution from the potential energy
to the second virial coefficient is
\begin{eqnarray}
B_2^{(\tilde w)}
& = &
\frac{V}{6}   \left\langle
 e^{\tilde w({\bf \Gamma})}\,
 {\bf \Gamma}^\dag \cdot \nabla_{\bf \Gamma} \tilde w({\bf \Gamma})
\right\rangle_{\rm id}
\nonumber \\ & = &
\frac{2\pi}{6}
 \int_{-\infty}^\infty {\rm d} p_z
\int_0^\infty {\rm d}{q} \, q^2
\int_{-1}^{1}  {\rm d} x\;
\nonumber \\ && \quad \times
\omega_p(p_z)
e^{\tilde w(p_{z},q_z,q)}
v^{\rm int}(p_{z},q_z,q)
\nonumber \\  && \quad \times
\Theta(\pi/2 - | \tilde w_{\rm i}|) .
\end{eqnarray}
Recall that $q_z = qx $,
$Z^{\rm id} = {V^2}/{2\Lambda^6}$,
and that
$ \omega_{p}(p_{z}) =
({\Lambda_\beta }/{h\sqrt{2}}) e^{-\beta  p_z ^2/4m } $.
Here we have restricted the integration
to the region where
the imaginary part of $\tilde w$ has magnitude less than $\pi/2$.
This restriction was not accounted for
in deriving the expression for the virial derivative,
since the consequent $\delta$-function contribution
is expected to be negligible.
Results with and without this restriction will be presented below.

We have a complex exponent
$\tilde w = \tilde w_{\rm r}  + {\rm i} \tilde w_{\rm i}  $,
and also its virial derivative
$v^{\rm int}=v^{\rm int}_{\rm r} + {\rm i} v^{\rm int}_{\rm i}$,
with the imaginary parts being odd in $p_z$.
These are the terms with $k$ odd in both cases.
In view of these  we have
\begin{eqnarray}
B_2^{(\tilde w)}
& = &
\frac{2\pi}{6}
 \int_{-\infty}^\infty {\rm d} p_z
\int_0^\infty {\rm d}{q} \, q^2
\int_{-1}^{1}  {\rm d} x\;
\nonumber \\  && \quad \times
\omega_p(p_z)
e^{\tilde w_{\rm r}} 
\left\{  v^{\rm int}_{\rm r} \cos \tilde w_{\rm i}
-  v^{\rm int}_{\rm i} \sin \tilde w_{\rm i} \right\}
\nonumber \\  && \quad \times
\Theta(\pi/2 - | \tilde w_{\rm i}|) .
\end{eqnarray}

From the computational point of view
since $\omega_p(p_z)$ is a Gaussian,
we can take $ p_z \alt 4 \sqrt{m/\beta}$.
The integrand is an even function of $p_z$,
and so we can save a factor of 2 by only integrating over the positive half.
We can terminate the position integral such that $q  \alt 3 \sigma$,
which is about the traditional cut-off for the Lennard-Jones potential.

Since in general
$ \int_{-\infty}^\infty {\rm d} p_z\, \omega_p(p_z) = 1$,
in the classical case this reduces to
\begin{eqnarray}
B_2^{\rm cl}
& = &
\frac{2\pi}{6}
\int_0^\infty {\rm d}{q} \, q^2
\int_{-1}^{1}  {\rm d} x\;
e^{-\beta u(q)} (-\beta) q u'(q)
\nonumber \\ & = &
\frac{-\beta}{6} \int {\rm d}{\bf q} \, e^{-\beta u(q)} q u'(q) .
\end{eqnarray}
This agrees with Attard (2026b Eq.~4.19)
and is the known result (Attard 2002, Pathria 1972).
This confirms that $\rho^2 = 2/V^2$
is correct for the present two-particle analysis.

%
\section{Computational Results} \label{Sec:Results}
\setcounter{equation}{0} \setcounter{subsubsection}{0}
\renewcommand{\theequation}{\arabic{section}.\arabic{equation}}
%

Previous calculations show that there is negligible difference
between the second virial coefficient for helium given by
the Hartree-Fock dispersion-B2 pair potential (Aziz 1992)
and by the Lennard-Jones 6-12 pair potential (Attard 2026b).
Therefore the present calculations are carried out with the latter,
$u(r) = 4 \varepsilon [(\sigma/r)^{12} - (\sigma/r)^{6} ]$,
with $\varepsilon = 10.22 k_{\rm B}$\,J
and $\sigma = 0.2556$\,nm (van Sciver 2012).
Obviously this makes the initial condition for the recursion relation
particularly simple, Eq.~(\ref{Eq:w1-LJ}).

Typical  quadrature parameters were
50 points for the momentum,
with $p_{\rm max} = 4 \sqrt{mk_{\rm B}T}$,
30 points for the angular integral,
and 60 points for the position integral,
with $q_{\rm min} = 0.5\sigma$, and  $q_{\rm max} = 4\sigma$.
These appear conservative and little change was observed
when they were varied significantly.
The power law expansion typically used
$j \in \{1,2,\ldots,6\}$, $k \in \{0,1,2,3\}$ $l \in \{0,1,\ldots,6\}$
and $n \in \{6,7,\ldots,12\}$.
Apart from $n$, which behaved poorly with $n_{\rm max} \ge 14$,
these again appeared to be conservative choices.
With this set 19 expansion coefficients were non-zero.
A full temperature curve with 75 points took several minutes to calculate
on a desk-top personal computer.

\begin{figure}[t]
\centerline{ \resizebox{8cm}{!}{ \includegraphics*{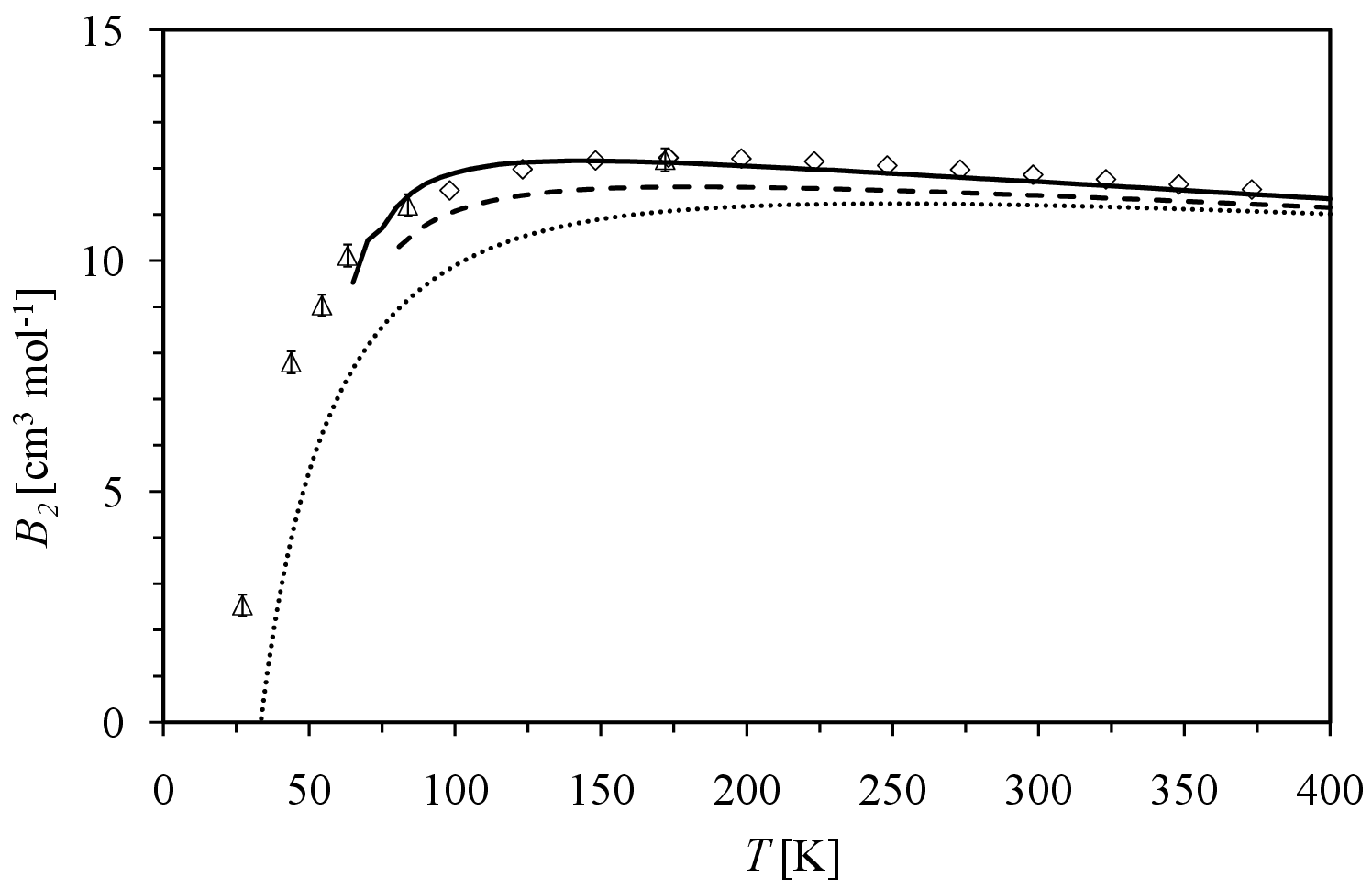} } }
\caption{\label{Fig:B2b}
Second virial coefficient of $^4$He.
The symbols are measured data
of Gammon (1976) (diamonds) 
and of Kemp \emph{et al.}\ (1986/87) (triangles). 
Using
the present power law series
for the pair commutation function with the Lennard-Jones potential,
the solid curve includes the weight factor
$\Theta(\pi/2 - |\tilde w_{\rm i}|)$
whereas the dashed curve doesn't.
The dotted curve is the classical result,
$\tilde w  = -\beta u$.}
\end{figure}

Figure~\ref{Fig:B2b} compares the measured
(Berry  1979, Gammon 1976, Kemp \emph{et al.}\ 1986/87)
and calculated second virial coefficient of $^4$He for $T \agt 30$\,K.
The measurement error is less than the size of the symbols,
and so these can be taken as reliable benchmarks.
That the second virial coefficient is positive for these temperatures
indicates that the pressure is dominated by the short-range repulsive part
of the pair potential
rather than by the  long-range attractive tail.

This last point is a little puzzling because
at low densities atoms are far apart on average,
and so their interaction might be expected to be dominated
by the attractive tail of the potential at long-range.
However one can understand the measurements as follows.
The force between the pair of atoms that gives the second virial coefficient
occurs when they are in the potential well.
(The non-force region beyond the range of the potential
is accounted for by the first virial coefficient.)
The attractive part is at larger separations,
$ 2^{1/6}\sigma < q \alt 3\sigma$,
and it has lower maximum magnitude than most of the steeply repulsive force
at closer separations.
At high temperatures the atoms are able to access
relatively deeply into the core,
and so the repulsive forces far outweigh the attractive forces
simply because the magnitude of the force is so much larger.
This is true even though the position configuration volume
for the attraction is larger than that for the repulsion.
As the temperature is lowered, less of the core is accessible,
and the relative advantage of the repulsive forces is reduced
because both their maximum value and the accessible volume are reduced.
Eventually, the greater volume of position configurations
for attraction, which is less sensitive to temperature,
overcomes the remaining advantage in magnitude of the repulsion
and the second virial coefficient turns negative.

Returning to Fig.~\ref{Fig:B2b},
all the calculations tend toward the measured data
as the temperature is increased.
The classical result (dotted curve) is surprisingly good,
although it does consistently underestimate the second virial coefficient.
The quantum calculation that places no restriction on
the imaginary part of the commutation function exponent (dashed curve)
lies between the classical result and the measured data.
The quantum calculation that restricts the allowed phase space points,
$|\tilde w_{\rm i}| < \pi/2$ (full curve),
is practically indistinguishable
from the laboratory measurements for temperatures $ T \ge 65$\,K.
This is fairly convincing evidence
that the restriction is both correct and necessary.

The kinetic energy contribution
to the second virial coefficient is generally
on the order of 1\% of the total.
This is not surprising as it takes the form of a fluctuation average.
At 70\,K it is about 3\%,
and at 65\,K it is about 10\% of the total,
further increasing thereafter.

Although  above 65\,K
the agreement between the full calculations and the measurements
is very good, it is not perfect.
The small discrepancies that are visible
might be attributed to either numerical error
or to the approximate nature of the Lennard-Jones pair potential
at short separations.

Some limited tests were performed
using the Hartree-Fock dispersion-B2 pair potential (Aziz 1992),
with the recursion relation starting from a fit of
$\sum_n w_{1,0,0,n} \tilde \beta \tilde q^{-n} $
to $-\beta  u(q)$, with $n \in \{6,7,\ldots,12\}$.
The fit was very accurate in the core region,
but underestimated the depth of the potential minimum
when $q_{\rm min} = 0.5\sigma$ was used,
and overestimated it using $q_{\rm min} = 0.6\sigma$.
In both cases the resultant second virial coefficient was much larger
than the measured  one, being 10--20\% too large at $T=400$\,K
and 30--50\% too large at $T=200$\,K,
and with no trend toward negative values at low temperatures.
The problem seems to be that the simplest fitting procedure
(minimizing the square of the error with weight $q^2$)
is dominated by the diverging potential in the core region.
In the region of the potential minimum
the fitted potential turns out to be in worse agreement
with the Aziz HFD-B2 potential
than is the Lennard-Jones potential itself.
Reducing the core contribution
by fitting instead only in the range $\tilde q \in [0.9,2.0]$
gave a relatively accurate minimum
and induced a maximum in $B_2$ at 70\,K,
but still with values 10--20\% too large.
Perhaps more coefficients, or a differentially weighted fit,
would improve matters.
However, in view of the good agreement with the measured data exhibited
in Fig.~\ref{Fig:B2b},
the Lennard-Jones pair potential is arguably good enough,
and it doesn't require any fitting parameters
beyond the established values of $\varepsilon$ and $\sigma$
(van Sciver 2012).

Below $T=65$\,K
the power law expansion for the commutation function exponent
begins to yield erratic results for the phase space weight.
At $T=60$\,K $B_2=-11.6$\,cc/mol,
and at $T=50$\,K, $B_2=1.5\times 10^4$\,cc/mol.
The failure of the power series beneath about 65\,K
was fairly consistent for different sets of numerical parameters.
Turning off the non-linear quadratic part of the recursion relation
made little difference for $T \agt 100$\,K,
where it reduced $B_2$ by about 2\%.
However, this gave $B_2=-18.7$\,cc/mol at $T=60$\,K
and $B_2=3.3\times 10^3$\,cc/mol at $T=50$\,K.


\begin{figure}[t]
\centerline{ \resizebox{8cm}{!}{ \includegraphics*{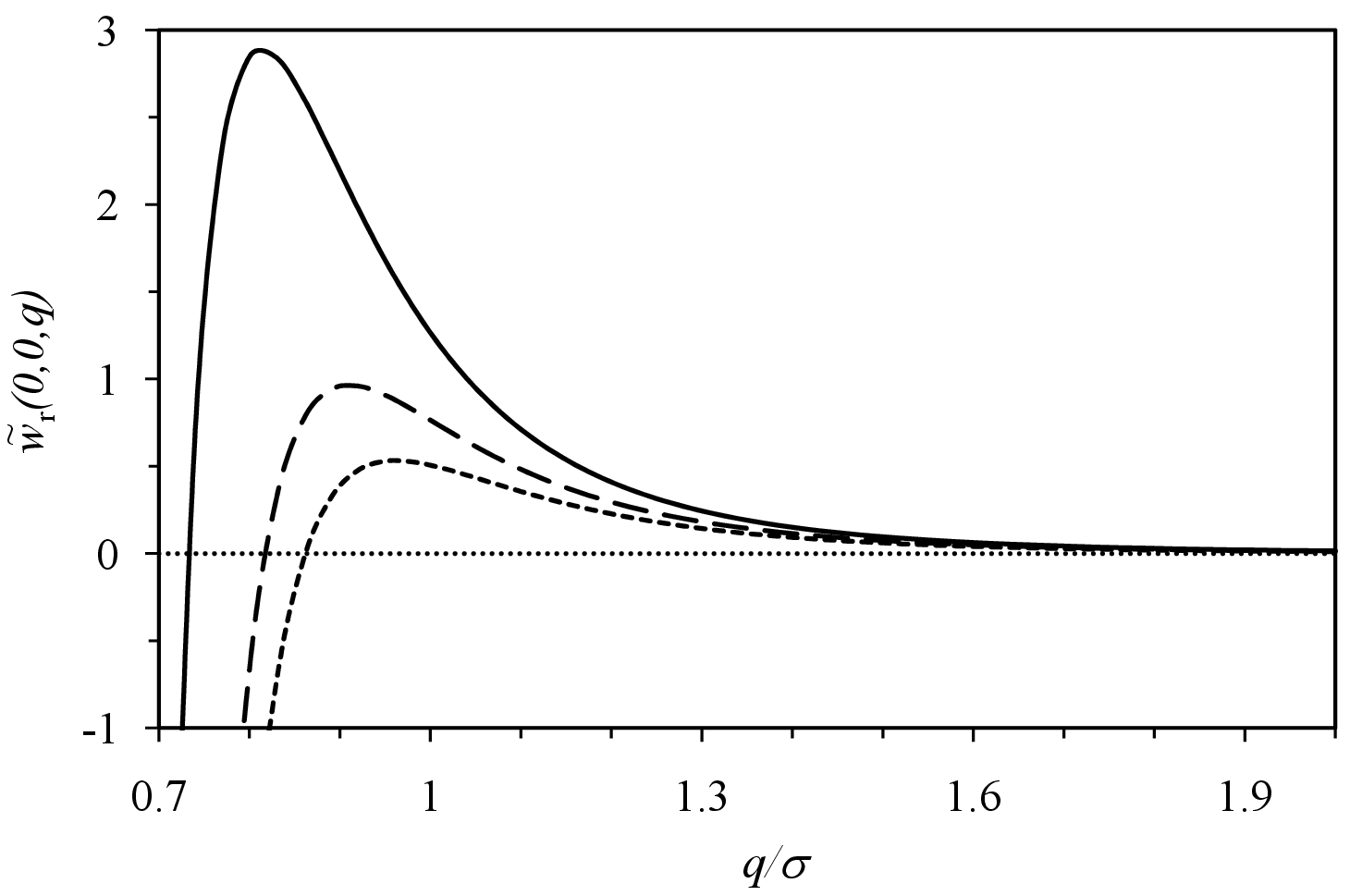} } }
\caption{\label{Fig:wr}
Real part of the pair weight exponent,
$\tilde w_{\rm r}(p_z,q_z,q)= w_{\rm r}(p_z,q_z,q) - \beta u(q)$,
on the line $p_z=q_z=0$
at $T=70$ (lower), 60 (middle), and 50\,K (upper).
}
\end{figure}

Figure~\ref{Fig:wr} shows
the real part of the  pair weight exponent,
$\tilde w_{\rm r}(p_z,q_z,q) $ on the line $p_z=q_z=0$,
where the imaginary part is zero anyway.
The exponent, which is essentially
the negative of an effective pair potential,
goes to zero at large separations
and to minus infinity at small.
The curves would appear unchanged if the non-linear term
was excluded from the recursion relation.
The peak of the exponent grows with decreasing temperature,
and it moves to smaller separations.
It first exceeds unity at about $T=60$\,K.
At $T=50$\,K it reaches 2.85 at $q=0.8\sigma$
(2.93 without the non-linear term).
This is more than fourteen times larger than
the maximum contribution from the Lennard-Jones potential itself,
which at $T=50$\,K is $\beta \varepsilon = 0.20$ at $q=2^{1/6}\sigma$.

%
\section{Conclusion} \label{Sec:Concl}
\setcounter{equation}{0} \setcounter{subsubsection}{0}
\renewcommand{\theequation}{\arabic{section}.\arabic{equation}}
%

As mentioned in the introduction,
the formulation of quantum statistical mechanics
in terms of configurations in classical phase space
requires a reliable means
of quantifying the Wigner-Kirkwood commutation function.
The power law expansion proposed
in the present paper is by far the most promising approach
that has been evaluated to date.
It yields very good agreement with measured data
for the second virial coefficient for $^4$He
for temperatures above about 65\,K.
This on its own is plausible proof
that the problem has been solved in the general case
in the high temperature, low density regime.

This, naturally, raises the question
of whether further progress can be made.
One limitation of the present approach
is that it is restricted to pair contributions to the commutation function.
In general terms three-body and higher effects
can be expected to become increasingly important as the density is increased.
One could build on the present approach
by developing power series for the many-body cases.
The feasibility of this is unclear,
although it is possible that the bulk of the correction
will come from three-body terms,
and that in any case one might need go no higher than four-body terms.
It should be pointed out that there is an arguable inconsistency
in pursuing many-body effects for the commutation function
while neglecting such effects for the interaction potential.
It may be that
the present pair commutation function may be sufficient even for liquids.

A second, perhaps more pressing, challenge
is to extend the treatment to lower temperatures.
The evidence so far gathered suggests
that the present failure for $T \alt 65$\,K,
and failures in previous approaches,
are related to the divergence of the pair potential
and its gradients in the core region,
which leads to the breakdown of the temperature expansion.

A significant result of the present paper
is the experimental confirmation
that the commutation function exponent should be restricted
to the primary branch in the positive real half of the complex plane,
$-\pi/2 < W_{\rm i} < \pi/2$.
With this restriction the results for the second virial coefficient
are in quantitative agreement with those measured,
but without the restriction there is measurable error.
This condition was previously proposed
on the basis of certain theoretic considerations
(the exponent is extensive,
and its imaginary part gives destructive oscillations
with infinitesimal period
for phase space points with $|W_{\rm i}|>\pi/2$)
(Attard 2025h).
At high temperatures in the classical regime,
$|W| \to  0$ as $T \to \infty$.
Attard (2025h) argued that the system could never cross the boundary
$|W_{\rm i}| = \pi/2$ as the temperature is lowered
because such boundary points have zero probability,
$\cos W_{\rm i} = 0$.
These theoretical arguments are bolstered
by the present comparison with experimental measurements.

\section*{References}


\begin{list}{}{\itemindent=-0.5cm \parsep=.5mm \itemsep=.5mm}

\item 
Allen M P and Tildesley D J 1987
\emph{Computer Simulation of Liquids}
(Oxford: Clarendon Press)

\item 
Attard P 2002
\emph{Thermodynamics and Statistical Mechanics:
Equilibrium by Entropy Maximisation}
(London: Academic)

\item 
Attard  P 2012
\emph{Non-equilibrium Thermodynamics and Statistical Mechanics:
Foundations and Applications}
(Oxford: Oxford University Press)

\item 
Attard P 2016b
Quantum statistical mechanics as an exact classical expansion with
results for Lennard-Jones helium
arXiv:1609.08178v3

\item
Attard P 2017
Quantum statistical mechanics results for argon, neon, and helium using
classical Monte Carlo
arXiv:1702.00096

\item 
Attard P 2018b
Quantum statistical mechanics in classical phase space. Expressions for
the multi-particle density, the average energy, and the virial pressure
arXiv:1811.00730

\item 
Attard P  2021
\emph{Quantum Statistical Mechanics in Classical Phase Space}
(Bristol: IOP Publishing)

\item 
Attard P 2025a
\emph{Understanding Bose-Einstein Condensation,
Superfluidity, and High Temperature Superconductivity}
(London: CRC Press)

\item 
Attard P 2025b
The molecular nature of superfluidity: Viscosity of helium from quantum
stochastic molecular dynamics simulations over real trajectories
arXiv:2409.19036v5

\item 
Attard P 2025g
Introduction to the modern theory of Bose-Einstein condensation,
superfluidity, and superconductivity
arXiv:2511.08953

\item
Attard P (2025h)  
Quantum Monte Carlo in classical phase space with the Wigner-Kirkwood
commutation function. Results for the saturation liquid density of $^4$He
arXiv:2512.09948v2.

\item 
Attard P 2026b
Gaussian Reformulation of the Feynman Path Integral
for Quantum Statistical Mechanics
with Results for the Second Virial Coefficient of $^4$He
arXiv:2607.16301

\item   
Aziz R A, Slaman M J,  Koide A, Allnatt  A R, and Meath W J 1992
Exchange-Coulomb potential energy curves for He-He,
and related physical properties
\emph{Mol.\ Phys.}\ {\bf 77} 321

\item
Berry K H 1979
NPL-75: A low temperature gas thermometry scale from 2.6\,K to 27.1\,K
\emph{Metrologia} {\bf 15} 89

\item 
Ceperley  D M  1995
Path integrals in the theory of condensed helium
\emph{Rev.\ Mod.\ Phys.}\ {\bf 67} 279

\item
Gammon B E 1976
The velocity of sound with derived state properties in helium
at -175 to 150 $^\circ$C with pressure to 150~atm.
\emph{J. Chem. Phys.}\ {\bf 64} 2556

\item
Kemp R C, Kemp W R G, and Besley L M 1986/87
A determination of thermodynamic temperatures and measurements
of the second virial coefficient of $^4$He between 13.81\,K and 287\,K
using a constant-volume gas thermometer
\emph{Metrologia} {\bf 23} 61

\item 
Kirkwood J G 1933
Quantum statistics of almost classical particles
\emph{Phys.\ Rev.}\ {\bf 44} 31

\item 
Pathria R K 1972
\emph{Statistical Mechanics} (Oxford: Pergamon Press)

\item 
van Sciver  S W 2012
\emph{Helium Cryogenics}
(New York: Springer 2nd edition)

\item 
Wigner E 1932
On the quantum correction for thermodynamic equilibrium
\emph{Phys.\ Rev.}\ {\bf 40} 749

\end{list}

\end{document}